\documentclass[twocolumn,english,aps,prl,floatfix,amssymb,groupedaddress]{revtex4}
\usepackage[T1]{fontenc}
\usepackage{amsmath}
\usepackage{amssymb}
\usepackage{graphicx}
\usepackage{lmodern}
\usepackage{wasysym}
\usepackage{pdfsync}
\usepackage[caption=false]{subfig}

\makeatletter
\@ifundefined{textcolor}{}
{%
 \definecolor{BLACK}{gray}{0}
 \definecolor{WHITE}{gray}{1}
 \definecolor{RED}{rgb}{1,0,0}
 \definecolor{GREEN}{rgb}{0,1,0}
 \definecolor{BLUE}{rgb}{0,0,1}
 \definecolor{CYAN}{cmyk}{1,0,0,0}
 \definecolor{MAGENTA}{cmyk}{0,1,0,0}
 \definecolor{YELLOW}{cmyk}{0,0,1,0}
}

\newcommand*\xbar[1]{%
   \hbox{%
     \vbox{%
       \hrule height 0.6pt 
       \kern0.4ex
       \hbox{%
         \kern-0.1em
         \ensuremath{#1}%
         \kern-0.1em
       }%
     }%
   }%
}

\usepackage{babel}
\usepackage[linkcolor=blue,urlcolor=blue,colorlinks,citecolor=blue]{hyperref}

\makeatother

\usepackage{babel}
\begin{document}

\title{Composite Weyl semimetal as a parent state for three dimensional topologically ordered phases}

\author{Eran Sagi}
\author{Ady Stern}
\author{David F. Mross}
\affiliation{Department of Condensed Matter Physics, Weizmann Institute of Science, Rehovot 76100, Israel}

\begin{abstract}
We introduce (3+1) dimensional models of short-range-interacting electrons that form a strongly correlated many-body state whose low-energy excitations are relativistic neutral fermions coupled to an emergent gauge field, $\text{QED}_{4}$. We discuss the properties of this critical state and its instabilities towards exotic phases such as a gapless `composite' Weyl semimetal and fully gapped topologically ordered phases that feature anyonic point-like as well as line-like excitations. These fractionalized phases describe electronic insulators. They may be further enriched by symmetries which results in the formation of non-trivial surface states.
\end{abstract}
\maketitle
{ \bf \emph{Introduction.}}~The notion of topological order as a characteristic of interacting quantum systems is by now widely recognized. Its hallmark is the existence of `fractional' quasiparticles---deconfined excitations carrying quantum numbers that cannot be obtained by any local combination of microscopic excitations~\cite{Wen2004}. Well-known examples of this occur in fractional quantum Hall states of electrons, which can host, e.g., charge-neutral fermions and fractionally charged anyons~\cite{Wen2004, Jain2007, DasSarma1996}. These exotic phases do not permit an easy description in terms of microscopic degrees of freedom and are more conveniently formulated in terms of `dual' variables. In the context of quantum Hall states, a non-local transformation known as flux attachment transforms electrons into composite fermions~\cite{Jain2007}. Simple mean-field phases of these composite fermions realize highly non-trivial topologically ordered phases of electrons, including non-Abelian ones such as the celebrated Moore-Read Pfaffian~\cite{Nayak2008}.

Recently, these ideas were adapted for symmetry protected topological (SPT) phases, where translating the action of local symmetries onto the non-local degrees of freedom adds an additional subtlety. This line of research led to discovery of a duality between a single (2+1) dimensional Dirac cone of weakly interacting electrons and the strongly interacting gauge theory $\text{QED}_{3}$ for a single flavor of dual fermions \cite{Son2015,Wang2015,Metlitski2016,diracduality.metlitski2,Mross2016}.  Subsequently, this fermionic duality was realized to be part of a larger `web of dualities' \cite{dualityweb.SeibergSenthilWangWitten, dualityweb.KarchTong, dualityweb.murugan, dualityweb.mirror.kachru1, dualityweb.mirror.kachru2, dualityweb.mross, dualityweb.lattice.ChenRaghu, dualityweb.goldman}. A natural application of the fermionic duality is the study of possible surface phases of 3D topological insulators (TIs), which feature a single electronic Dirac cone and can thus equivalently be described as $\text{QED}_{3}$ for dual fermions. This dual description is a natural `parent state' of various topologically ordered phases, such as the `T-Pfaffian' which arises when dual fermion form a time-reversal invariant BCS superconductor~\cite{Bonderson2013, Wang2013, Chen2014, Metlitski2015}.

In this work, we leverage these insights to develop a strongly correlated parent state in (3+1) dimensions, where emergent fermions couple to
a dynamical gauge field, similar to the (3+1) dimensional theory of quantum electrodynamics---$\text{QED}_{4}$. This is achieved by combining the (2+1) dimensional duality with the mechanism of \textit{spontaneous inter-layer coherence} of the emergent fermions \cite{JasonGilInterlayer, You2017}, which deconfines them in the third dimension. Consequently, this emergent QED$_4$ only arises as a consequence of strong interactions, unlike the QED$_3$ that constitutes an equivalent (dual) description of free Dirac fermions.

Despite this difference, pairing of emergent fermions has an analogous effect to that occurring in $(2+1)$ dimensions. In both cases, the emergent gauge field acquires a gap, leading to various fractionalized phases. Specifically, we will discuss gapless composite Weyl semimetals, gapped topologically ordered phases, and symmetry-enriched topological phases (SETs).  The first of these---the `composite' Weyl semimetal---is an electric insulator that features Majorana-Weyl nodes of emergent fermions in the bulk. The composite Weyl semimetal phase should be contrasted with the fractional chiral metal proposed in Ref. \cite{Meng2016} and the anomalous semimetal of Ref.~\cite{TeoWeyl2017}, which are not insulating. Merging the Majorana-Weyl nodes results in a fully gapped phase  with fractional point-like and line-like excitations. In fact, the resulting phase is the 3D analog of the toric code \cite{Kitaev2003}. Finally, by carefully tracing the action of electronic symmetries onto these degrees of freedom, we can use the understanding of \textit{free fermion} topological superconductors to distinguish SETs, which may carry symmetry protected fractional surface states.

\begin{figure}
\includegraphics[width=\columnwidth]{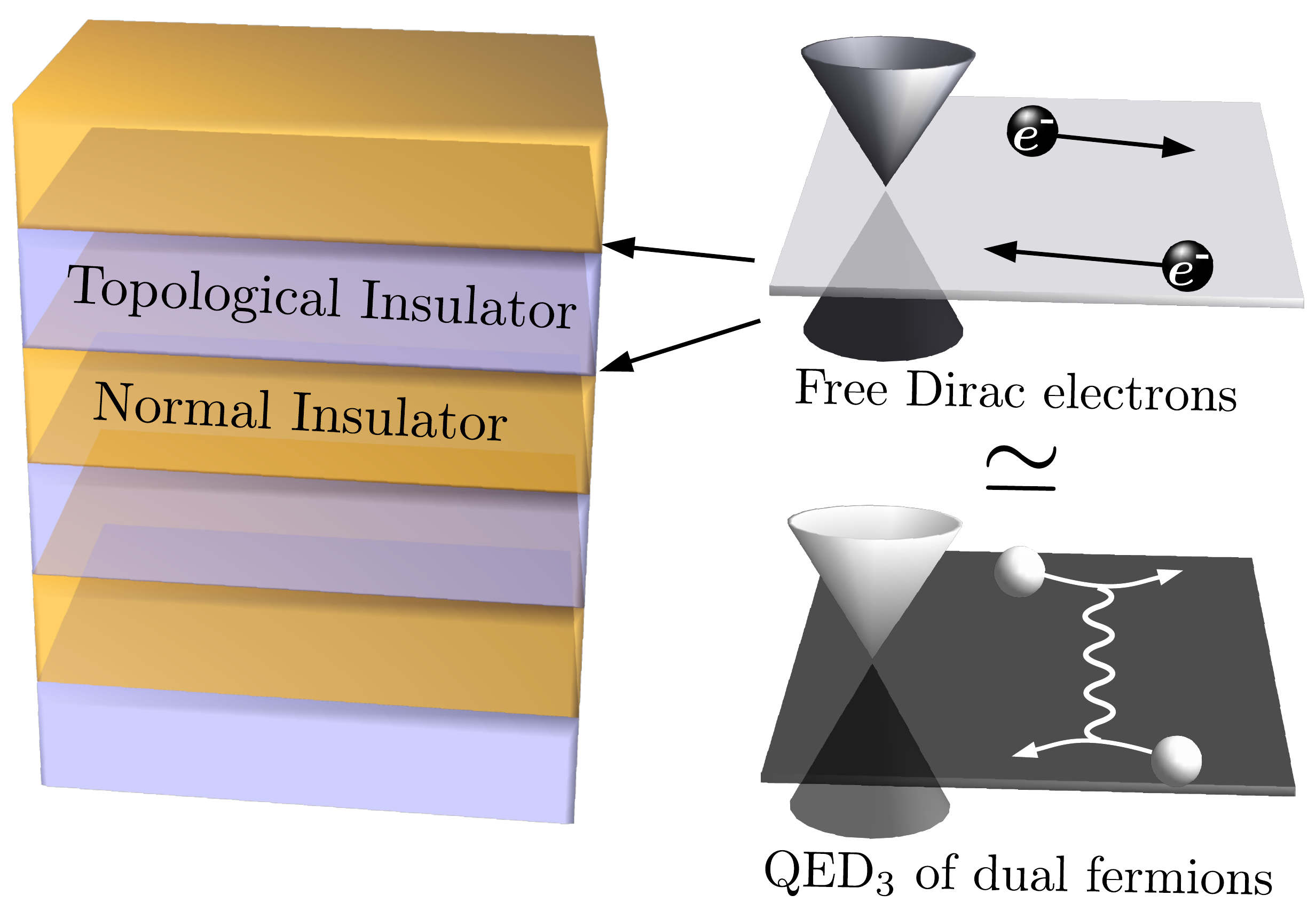}
\caption{ A schematic depiction of the superlattice we study.
The system is composed of alternating topological and normal insulators.
 Each interface contains a single Dirac cone, known to be dual to a
$\text{QED}_{3}$ theory describing neutral fermions coupled to an
emergent gauge field. We demonstrate that interactions can deconfine
the neutral fermions, leading to a $\text{QED}_{4}$ theory, from
which non-trivial three-dimensional phases descend.}
\label{fig:stack}
\end{figure}

{ \bf \emph{The model.}}~ Our model consists of alternating
thin layers of topological and normal insulators, as shown in Fig.~\ref{fig:stack}. Such a configuration was previously used in Ref. \cite{Burkov2011} to construct a Weyl semimetal phase \cite{ShuichiMurakami2007,Wan2011,Hasan2017,Hosur2013,Xu2015b,Lv2015,Huang2015,Weng2015,Xu2015a,Zheng2016,Yang2015}. The interfaces between adjacent layers are enumerated
by the integer index $z$. In the simplest case, each interface contains
a single Dirac cone, described by the continuum action
\begin{align}
{\cal S} & =i\sum_{z}\int d^{2}xdt \ \xbar{\Psi}_{z}\gamma_{\mu z}\left(\partial^{\mu}-iA^\mu\right)\Psi_{z}~,\label{eq:Dirac}
\end{align}
where $\Psi_{z}=\begin{pmatrix}\psi_{z\uparrow} & \psi_{z\downarrow}\end{pmatrix}^{T}$
are Dirac electrons; $A^\mu$ are components of electromagnetic potential with space-time index $\mu=0,1,2$; $\gamma$-matrices are represented as $\gamma_{0z}=\sigma_{1}$,
$\gamma_{1z}=-i(-1)^{z}\sigma_{2}$,$\gamma_{2z}=i(-1)^{z}\sigma_{3}$; and $\xbar{\Psi}=\Psi^\dagger \gamma_{0z}$. The theory possesses anti-unitary time reversal and particle-hole
symmetries of the form $\mathcal{T}\Psi\mathcal{T}^{-1}=i\sigma_{2}\Psi$
and $\mathcal{C}\Psi\mathcal{C}^{-1}=\sigma_{1}\Psi^{\dagger}$, which are broken by non-zero magnetic flux $\vec \nabla\times \vec A$ and chemical potential $A^0$, respectively.

Each (2+1) dimensional Dirac theory can be equivalently described by a dual $\text{QED}_{3}$  \cite{Son2015,Wang2015,Metlitski2016,diracduality.metlitski2,Mross2016}, with dual Dirac fermions $\widetilde{\Psi}_z = \begin{pmatrix}\tilde{\psi}_{z\uparrow} & \tilde{\psi}_{z\downarrow}\end{pmatrix}^{T}$ coupled to an \textit{emergent} $U(1)$ gauge field
$a^{\mu}$. The dual action for our multi-layer system is
given by ${\cal S}_{\text{dual}} = \sum_z \int d^{3}x {\cal L}_\text{dual,$z$}$
\begin{align}
{\cal L}_\text{dual,$z$}&=i\xbar{\widetilde{\Psi}}_{z}\gamma_{\mu z}\left(\partial^{\mu}-ia_{z}^{\mu}\right)\widetilde{\Psi}_{z}+ \frac{a_{\mu,z}\epsilon_{\mu\nu\kappa}\partial_\nu A_{\kappa,z} }{4\pi}+ \ldots~, \label{Eq:dual action}
\end{align}
where the ellipsis denotes `generic' terms such as a Maxwell action for $a$ and short-range interactions between the dual fermions.
Importantly, the roles of particle-hole and
time reversal symmetries are interchanged in the dual formulation: $\mathcal{T}\widetilde{\Psi}\mathcal{T}^{-1}=\sigma_{1}\widetilde{\Psi}^{\dagger}$
and $\mathcal{C}\widetilde{\Psi}\mathcal{C}^{-1}=i\sigma_{2}\widetilde{\Psi}$. The single electron of the original theory corresponds to a $4\pi$ monopole in the gauge field, which in turn binds two fermionic zero modes \cite{dualityweb.SeibergSenthilWangWitten}, one of which is occupied. In the continuum, the $4\pi$ monopole therefore binds a dual fermion, in analogy to the familiar attachment of two flux quanta in the conventional composite-fermion approaches.

Next, we couple the layers to form 3D phases, similar to the layer construction presented in Ref.~\cite{Jian2014}. The simplest such coupling, direct inter-layer tunneling of electrons, involves the insertion of $4\pi$ monopoles in the various layers, and therefore confines the gauge field, leading to free fermions phases. We initially tune such couplings to zero and will allow them back at a later stage. Non-trivial phases could easily arise if dual fermions (instead of electrons) were able to tunnel between different layers. However, these are not local excitations and there is no microscopic process that allows them to transfer between layers, i.e.,  dual fermions are confined in the $z$ direction. They may, however, become liberated when interactions between different layers generate strong inter-layer correlations.

{ \bf \emph{Inter-layer coherence.}}~To illustrate this mechanism we follow Ref.~\cite{JasonGilInterlayer} and consider a simple density-density interaction term of the form ${\cal S}_\text{\text{int}}=\sum_{z}\int d^{2}x dt {\cal L}_{\text{int},z}$ with ${\cal L}_{\text{int},z}=u\sum_{\alpha=\uparrow,\downarrow}\widetilde{\Psi}_{z,\alpha}^{\dagger}\widetilde{\Psi}_{z,\alpha}\widetilde{\Psi}_{z+1,\alpha}^{\dagger}\widetilde{\Psi}_{z+1,\alpha}$. Next, we use (dynamical) Hubbard-Stratonovich fields $\chi_\alpha$ to decouple the interaction as
\begin{equation}
{\cal L}_{\text{int},z}=\sum_{\alpha=\uparrow,\downarrow}\left(\chi_{z,\alpha}\widetilde{\Psi}_{z,\alpha}^{\dagger}\widetilde{\Psi}_{z+1,\alpha}+\text{H.c.}\right)+\frac{\left|\chi_{z,\alpha}\right|^{2}}{u}~.\label{HS trans}
\end{equation}
Notice that under a gauge transformation $\widetilde{\Psi}_{z}\rightarrow\widetilde{\Psi}_{z}e^{i\phi_{z}}$
and $a_{j}^{\mu}\rightarrow a_{z}^{\mu}+\partial_{\mu}\phi_{z}$.
Imposing gauge invariance then requires that $\chi_{z,\alpha}$
transforms as $\chi_{z,\alpha}\rightarrow\chi_{z,\alpha}e^{i\left(\phi_{z}-\phi_{z+1}\right)}.$
It is convenient to write $\chi_{z,\alpha}=\chi_{z,\alpha}^{0}e^{-i a_{z}^{3}}$ such that $a_{z}^{3}\rightarrow a_{z}^{3}+\phi_{z+1}-\phi_{z}$
under gauge transformations and the magnitude $\chi_{z,\alpha}^{0} =|\chi_{z,\alpha}|$ is gauge invariant. For sufficiently strong interactions the system may enter a phase with $\langle \chi_{z,\alpha}^{0}\rangle = \chi_0 \neq 0$ where fluctuations of the magnitude $\left|\chi_{z,\alpha}^{0}\right|$ are massive while phase fluctuations remain gapless.  The effective Lagrangian of Eq.~\eqref{HS trans} then becomes
\begin{equation}
{\cal L}_{\text{int},z}=\chi_0\left(e^{-i a_{z}^{3}}\widetilde{\Psi}_{z}^{\dagger}\widetilde{\Psi}_{z+1}+\text{H.c.}\right)+\frac{1}{\kappa}(\partial_\mu a^3_z)^2~,\label{eq: HS mean field}
\end{equation}
where the precise value of the coupling $\kappa$ depends on details and is unimportant here. The density-density interaction term thus induces
coherence between the layers and generates dual fermion tunneling. In the Appendix, we demonstrate this explicitly in a tractable coupled-wire model, similar to the one used in Ref. \cite{TeoWeyl2017} to construct strongly correlated insulators and anomalous metals.
We note that since we no longer have independent gauge symmetries for each layer, the gauge-field should be thought of as a 3D gauge field, with $a^3$ defined above acting as its $z$-component.

It is instructive to first analyze the system in a mean-field approximation where all four components of $a$ vanish. In this case, the model is non-interacting and the problem is reduced to studying the band structure of the dual fermions. Since the band structure is equivalent to the one studied in Ref.~\cite{Burkov2011} (at the time-reversal and inversion symmetric point), we expect to have a $(3+1)$-dimensional Dirac theory at low energies. Therefore, reintroducing the gauge degrees of freedom, we expect to obtain the $(3+1)$-dimensional theory of Dirac fermions coupled to a $U(1)$ gauge field---$\text{QED}_4$---described by the action
\begin{align}
{\cal S}_\text{QED$_4$} & =i\int d^{3}xdt \ \xbar{\widetilde{\Psi}}\hspace{0.2mm} ^\prime\Gamma_{\mu}\left(\partial^{\mu}-ia^{\mu}\right)\widetilde{\Psi}' + \ldots~,\label{eq:Dirac action}
\end{align}
where $\mu=0,\ldots,3$ is the spacetime index, $\widetilde{\Psi}'$ is a 4-dimensional Dirac spinor, and $\Gamma_\mu$ are $4\times4$ matrices satisfying the Clifford algebra. Notice that despite the isotropic appearance of Eq.~\eqref{eq:Dirac action} the physical properties of this system are highly anisotropic. The coupling of the physical electromagnetic field $A^\mu$ is the same as in Eq.~\eqref{Eq:dual action} and $A^3$ completely decouples from ${\cal S}_\text{QED$_4$}$. A detailed derivation of this low-energy description is provided in the Appendix. In what follows we turn
to investigate the properties of this non-trivial fixed point and
its instabilities.

{ \bf \emph{Inter-layer tunneling of electrons.}}~While the $\text{QED}_{4}$ state is not expected to describe a stable state, it is nevertheless natural to ask about its fate upon including tunneling of \textit{electrons} between neighboring layers. On general grounds, we expect that building up the correlations required for inter-layer tunneling of dual fermions will simultaneously suppress the electronic Green function. On the other hand, $\text{QED}_{4}$ is weakly coupled at long distances and the Green function of dual fermions is essentially free. Following this reasoning, the dual fermion has a smaller scaling dimension than the electron and we expect that the inter-layer coherent state is unstable against weak electron tunneling, at least in some parameter range.

To support this expectation, recall that an operator that creates physical charge $n e$ corresponds to a $4\pi n$ monopole insertion $M_n$ in the gauge theory \cite{Wang2015,Metlitski2016,diracduality.metlitski2,dualityweb.SeibergSenthilWangWitten, dualityweb.KarchTong}. Making use of the fact that $\text{QED}_{4}$ is free in the infrared, we find the monopole propator
\begin{align}\left\langle M_n(x)M_n^{\dagger}(0)\right\rangle \sim\left|x\right|^{-q^{2} n^2/4\pi^{2}}.\end{align} (See Ref.~\cite{Marino1994} and the Appendix for an alternative derivation.) The value of the non-universal number $q$ is set by the short-range interactions of the microscopic electrons. We take this result to indicate that the scaling dimension of the physical electron in the emergent $\text{QED}_{4}$ varies continuously, analogous to the case of Luttinger liquids in $1+1$ dimensions. Hence, we expect to find a range of parameters for which inter-layer tunneling of electrons is irrelevant.

The electronic conductivity of the $\text{QED}_{4}$ state can easily
be computed. In the absence of electronic tunneling between layers,
charge in each layer is conserved and the action is independent
of $A_{3}$. Again assuming that dual fermions decouple from the gauge field, we integrate out $a$ to obtain the effective action $\propto\sum_{\alpha=0}^{2}\left(A_{\alpha}\right)^{2}$ for the electromagnetic vector potential. It follows that the $\text{QED}_{4}$ state is superconducting in each plane and insulating in the $z$ direction. Including irrelevant electron tunneling allows inter-layer currents and thus reintroduces $A_3$. The conductivity in the $z$ direction then vanishes as $\omega \rightarrow 0$ with a non-universal power law that is again related to the coupling $q$.

{ \bf \emph{Composite Weyl semimetal.}}~The $\text{QED}_{4}$ fixed point
described above holds only at a fine-tuned point in parameter space.
We now turn to study the phases resulting from perturbing this theory.
First, we introduce the perturbation
\begin{equation}
{\cal S}_{m}=m\sum_{z}\int d^{2}x dt\ \xbar{\widetilde{\Psi}}\hspace{0.2mm} ^\prime_z\widetilde{\Psi}_{z}=m\sum_{z}\int d^{2}x dt\ {\xbar\Psi}_{z}{\Psi}_{z}~,\label{eq:time reversal breaking}
\end{equation}
which acts as a mass term in the decoupled layer limit $\chi_0=0$. This term matches the time-reversal breaking terms introduced in Ref.~\cite{Burkov2011}, which was shown to stabilize a Weyl semimetal phase as long as $m/\chi_0$ is small enough. We therefore expect to obtain a low-energy theory consisting of two Weyl fermions coupled to a gauge field. As we demonstrate in the Appendix, this is indeed the case, and the low energy action is given by
\begin{equation}
{\cal S}_{\text{Weyl}}=i\sum_{\beta=1}^2 \int d^{3}xdt\ \widetilde{\Psi}_{\beta}^{\dagger}\sigma_{\mu}^{\beta}\left(\partial^{\mu}-ia_{j}^{\mu}\right)\widetilde{\Psi}_{\beta}~,\label{Eq:Weyl action}
\end{equation}
where $\beta$ enumerates the two Weyl cones, the $\Psi_\beta$-fields are Weyl spinors, $\sigma_{\mu}^\beta=\sigma_\mu$ for $\mu = 0,\ldots,2$ and $\sigma_{3}^\beta = (-1)^\beta\sigma_3$.
Notice that while the two Weyl theories are technically equivalent
to a single Dirac theory, they are associated with degrees of freedom
located at different momenta $k_z=\pm \cos^{-1}\left(\frac{\left|m\right|}{2\chi_{0}}\right)$, prompting us to
regard them as separate theories.

\begin{figure}
\includegraphics[width=\columnwidth]{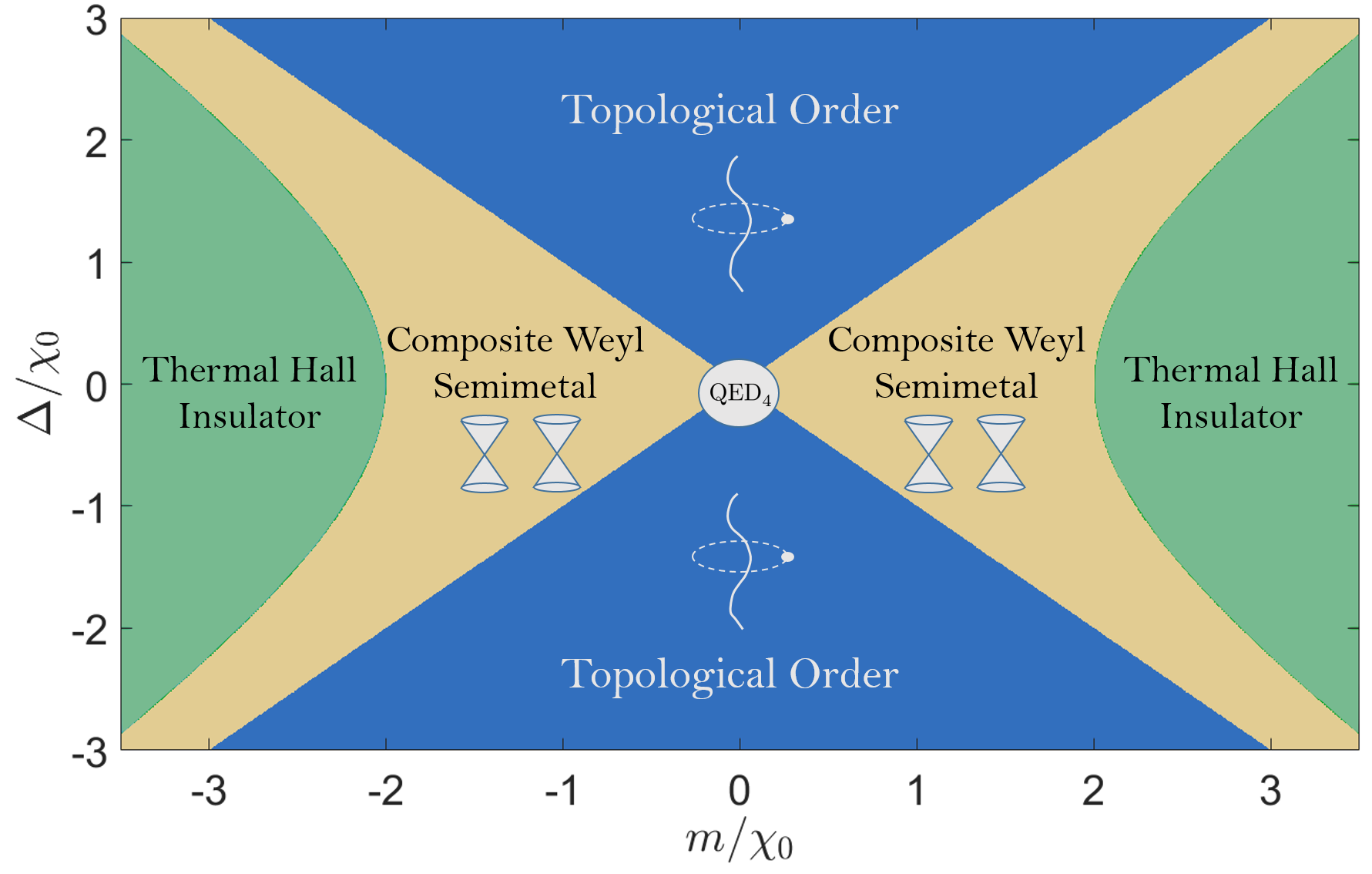}

\caption{\label{fig:phase diagram}The phase diagram describing the properties
of the system as a function of the mass $m$ and the amplitude $\Delta$
of the dual pairing terms. We show that the $\text{QED}_{4}$ theory
serves as a parent state for the composite Weyl semimetal state and topologically
ordered gapped phases. If the mass is large enough, the system becomes a thermal Hall insulator, with the dual fermions forming a band insulator which is adiabatically connectable to a stack of Chern insulators. }
\end{figure}

Furthermore, notice that the dual fermion number is not a microscopically
conserved quantity, meaning that pairing terms are naturally generated.
The simplest of these takes the form
\begin{equation}
S_{\Delta}=\Delta\sum_{z}\int d^{2}xdt\left(\widetilde\Psi_z i\sigma_2 \widetilde\Psi_z+\text{H.c.}\right)~.\label{eq:pairing}
\end{equation}
Such a term has two important effects: First, it provides a Higgs mass to the gauge
field. In addition, the low energy Weyl fermions are shifted to $k_{z}=\pm\cos^{-1}\left(\frac{\sqrt{m^{2}-\Delta^{2}}}{2\chi_0}\right).$
This means that the composite Weyl semimetal phase survives as long as
$\left|\Delta\right|\leq\left|m\right|$ and $m^{2}\leq4\chi_{0}^{2}+\Delta^{2}$
(see Fig.~\ref{fig:phase diagram} for the phase diagram). We note that in the more generic case, where $\Delta$ is not the same in all layers, each Weyl cone splits into two Majorana-Weyl cones \cite{Meng2012}. In such cases, one can access an additional Majorana-Weyl semimetal phase, where two of the four cones couple and become massive, resulting in only two gapless Majorana-Weyl cones.

Once the dynamical gauge field $\vec a$ is gapped, the electric current $\vec{J}\propto \vec{\nabla}\times \vec{a}$ is suppressed, and the system becomes electrically insulating.
However, there are still gapless neutral excitations in the bulk whose contribution to the thermal transport is the same as for the original electrons. We would like to compare this composite Weyl semimetal phase to the `composite Dirac liquid' (CDL), proposed to arise on the surface of a 3D topological insulator \cite{Mross2015}. The CDL state arises from QED$_3$ when the gauge degrees of freedom are gapped by the formation of a dual-fermion superconductor, leaving behind a Dirac cone of neutral fermions. The Dirac point is not protected by any symmetries and the CDL thus requires a certain amount of fine tuning. In contrast, the composite Weyl semimetal represents a stable phase of matter.

{ \bf \emph{Topological ordered and symmetry enriched phases.}}~Starting with the composite Weyl semimetal, one can readily access two distinct fully gapped phases. The first phase, represented by the blue regions in Fig.~\ref{fig:phase diagram}, is obtained when the two Weyl nodes meet on the edge of the Brillouin zone and annihilate. The second phase, depicted by the green regions in Fig.~\ref{fig:phase diagram}, is obtained when the two Weyl points coincide at the origin. As we demonstrate in the Appendix, the above indicates that the two phases are distinguished by their topological properties, with one of them having a completely trivial
band structure and the other forming a 3D stacked quantum Hall state of dual fermions, referred to as a thermal Hall insulator.

Regardless of the properties of band structure, when the system
exhibits a pairing gap, the gapped excitations consist
of the dual fermions and vortex lines. As usual, when a dual fermion
encircles a vortex line, it acquires a phase of $\pi$. Therefore, our system hosts deconfined point- and line-like excitations,
with non-trivial mutual statistics, i.e., it exhibits topological order.
In fact, it realizes the three-dimensional incarnation of
the toric code \cite{Kitaev2003}, where the two types of point-particles
exhibit the same mutual statistics.

The topological order is correct for any gapped superconducting state.
It is therefore natural to ask whether we can also stabilize phases
associated with strong topological band indices and non-trivial surface
states.
We note that the only topological class in 3D allowing for topologically
non-trivial superconducting phases is class DIII \cite{Schnyder2008,Kitaev2009},
with time reversal symmetry satisfying $\mathcal{T}^{2}=-1$. In the non-interacting limit, this
class exhibits infinitely many distinct topological phases, each of
which is described by an integer valued topological invariant. Interactions reduce the number of topologically distinct phases, leading to a $\mathbb{Z}_{16}$ classification \cite{Fidkowski2013, Metlitski2014,Senthil2015,Wang2015a}. The topologically ordered state discussed above has a ground state degeneracy of eight
on a three-dimensional torus. This degeneracy is consistent with time-reversal symmetry, hinting that it might indeed be possible to enrich it with a DIII topological band index.

In our case, the original time-reversal symmetry acts as an effective
anti-unitary particle-hole (or chirality) symmetry in the dual formulation. The presence of pairing terms leads to an additional unitary particle-hole symmetry. Multiplying the two symmetries, we obtain an anti-unitary time-reversal symmetry $\widetilde{\mathcal{T}}$ satisfying $\widetilde{\mathcal{T}}^{2}=-1$. Thus, as long as the physical time-reversal symmetry is preserved, the dual action indeed belongs to class DIII, and symmetry enriched phases can arise.

If the system is in a phase with a non-zero strong invariant, the
surface hosts an integer number of symmetry protected Majorana cones,
comprised of dual fermions.

{ \bf \emph{$\text{QED}_{\boldsymbol{4}}$ at finite density.}}~Finally, we want to briefly mention the possibility of `doping' the emergent fermions to a non-zero density by applying a magnetic field along the stacking direction. The (2+1)d duality maps physical magnetic flux onto dual-fermion density and thus each layer features a Fermi surface of dual fermions---essentially the composite Fermi liquid (CFL) that arises in the half-filled Landau level. Upon forming inter-layer coherence, these two-dimensional CFLs turn into a single three dimensional Fermi surface which may be more stable against various instabilities than the Dirac theory. A closely related three-dimensional state of composite Fermions (but without a Dirac dispersion) was envisaged in Ref.~\cite{JasonGilInterlayer} as a possible description of layered semimetals such as graphite under a strong magnetic field. Within our model, such a state corresponds to introducing a gap to the emergent Dirac fermions that is smaller than their chemical potential.

{ \bf \emph{Conclusions.}} We have introduced a strongly correlated critical theory that provides easy access to exotic three-dimensional phases such as gapless insulators and Abelian topological orders. To capture this parent state we used a known two-dimensional duality relation and extended it into the third dimension via `spontaneous inter-layer coherence'. We describe this mechanism both heuristically within a transparent but uncontrolled mean-field formulation and via an explicit coupled-wire construction. The latter moreover allows us to provide concrete Hamiltonians of microscopic electrons that realize these phases. Generalizing our approach to some of the other, recently discovered (2+1)d dualities could yield additional (3+1)d states that may realize different, and potentially even more exotic phases. A particularly interesting case would be a non-Abelian three dimensional topological order that could realize non-trivial `three-loop braiding' \cite{Wang2014, Jian2014}.

\begin{acknowledgments}
{ \bf \emph{Acknowledgments.}}
We are indebted to Jason Alicea and Yuval Oreg for useful discussions. This work was supported by the
Israel Science Foundation (DFM and AS); the Minerva
foundation with funding from the Federal German Ministry
for Education and Research (DFM); the Binational
Science Foundation (DFM); the European Research
Council under the European Communitys Seventh Framework
Program (FP7/2007-2013)/ERC Project MUNATOP (AS); Microsoft Station
Q (AS); the DFG (CRC/Transregio 183, EI 519/7-1) (AS); the Adams Fellowship Program of the Israel
Academy of Sciences and Humanities (ES).

\end{acknowledgments}
\bibliography{ref}

\newpage
\onecolumngrid
\appendix
\section{\large Appendix}

\section{Inter-layer coherence through a tractable model}
\label{Appendix:coupled-wire}
\subsection{The coupled-wire model}

When constructing the $\text{QED}_{4}$ state, we relied on the formation
of spontaneous inter-layer coherence, which in turn allowed us to write
inter-layer tunneling of dual fermions. In this section, we go beyond
the mean-field approach presented in the main text and demonstrate
the formation of inter-layer coherence through a tractable coupled-wire
model.

\begin{figure}
\subfloat[\label{fig:CDL}]{\includegraphics[scale=0.23]{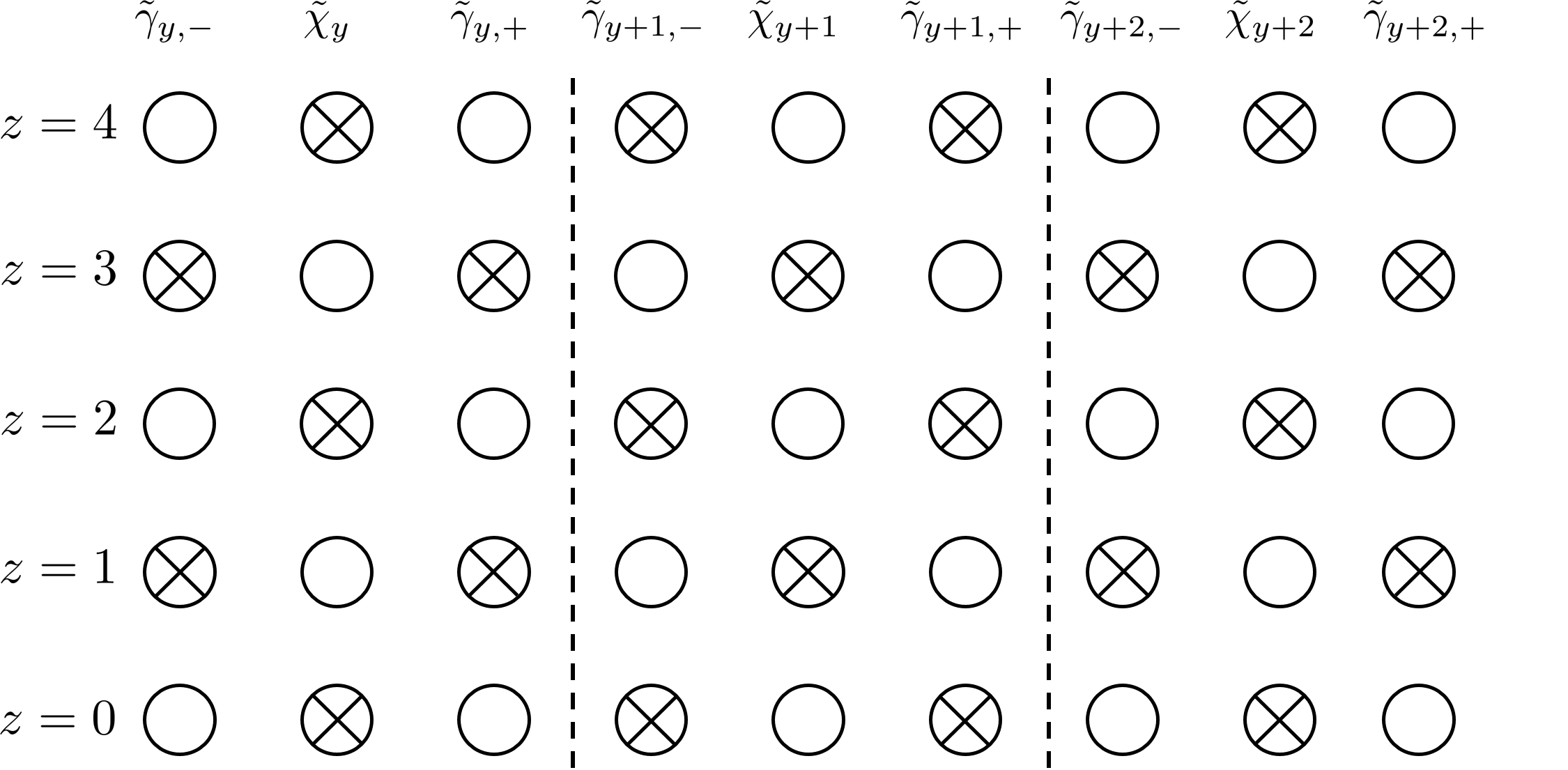}

}\subfloat[\label{fig:CDLcoherence}]{\includegraphics[scale=0.23]{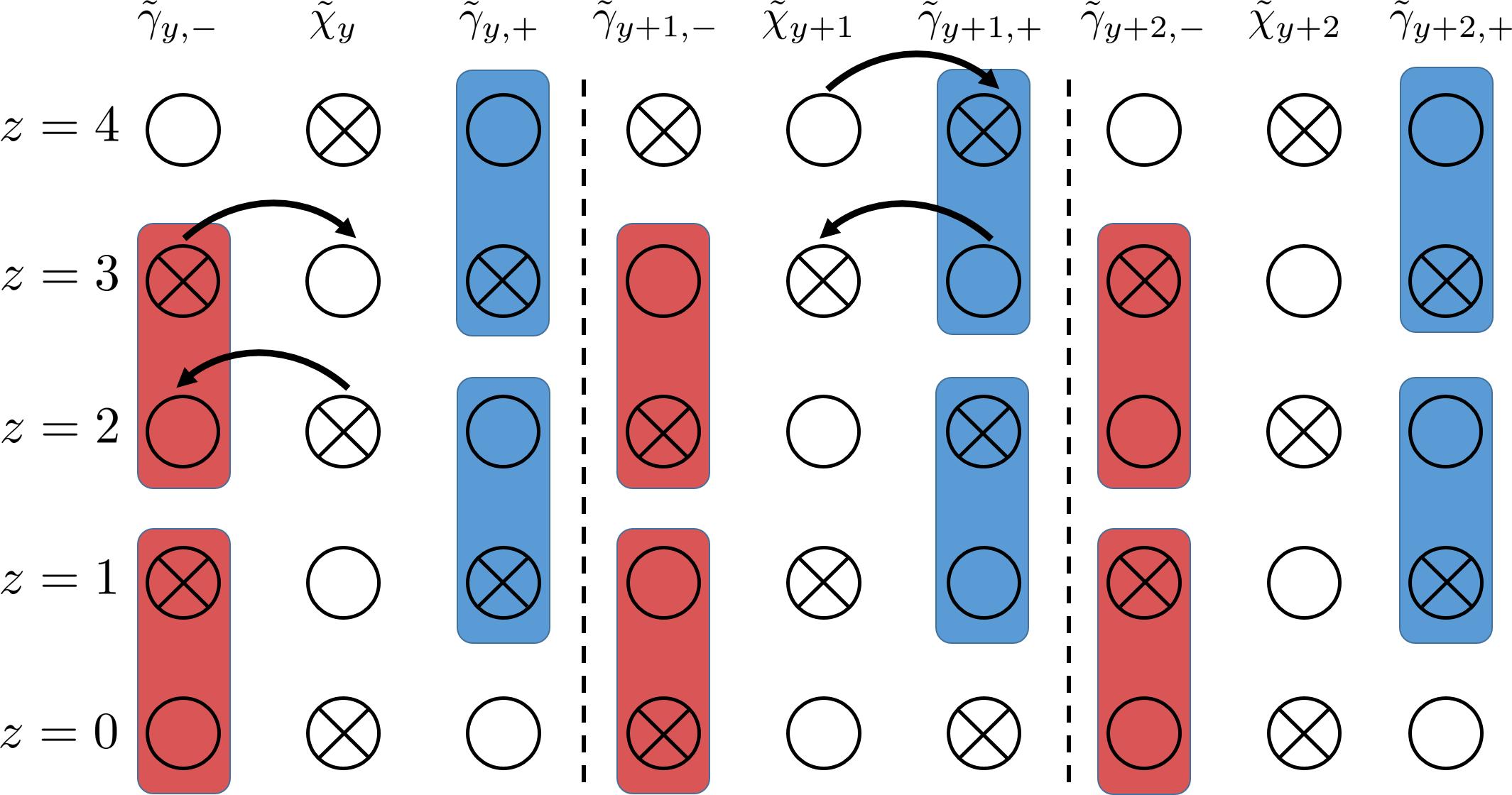}

}\caption{(a) A schematic depiction of our model, where the dual Dirac fermions
in each layer are represented by an array of coupled wires. Here,
the symbol $\fullmoon$ denotes a left mover, while $\otimes$ denotes
a right mover. (b) Coherence between different layers is induced when
$\left\langle \tilde{\gamma}_{yz+}^{\dagger}\tilde{\gamma}_{y,z+1,+}\right\rangle $
(blue) and $\left\langle \tilde{\gamma}_{yz-}^{\dagger}\tilde{\gamma}_{y,z+1,-}\right\rangle $
(red) acquire expectation values for odd and even $z$, respectively.
Then, four-fermion terms of the form shown by the arrows yield tunneling
of $\tilde{\chi}$ across layers.}
\end{figure}

The interfaces in our superlattice structure form two-dimensional planes, and will henceforth be referred to as layers.
In what follows, each layer is modeled
by an array of coupled wires \cite{Mross2016}. Each wire hosts a
chiral electronic mode, allowed to tunnel between adjacent wires. The interfaces form two-dimensional planes, and will henceforth be referred to as layers.
The wires in each layer are enumerated by the index $y$, the interfaces
are enumerated by the index $z$ (see Fig.~\ref{fig:CDL}), and the
electronic annihilation operators are given by $\chi_{y,z}$. The
Hamiltonian of the decoupled chiral wires is given by
\begin{equation}
H_{\text{wires}}=-iv\sum_{yz}(-1)^{y+z}\int dx\chi_{y,z}^{\dagger}\partial_{x}\chi_{y,z},\label{eq:wires-1}
\end{equation}
and the tunneling terms within each layer take the form

\begin{equation}
H_{\text{hop}}=t\sum_{yz}(-1)^{y+z}\int dx\left(\chi_{y,z}^{\dagger}\chi_{y+1,z}+\text{H.c.}\right).\label{coupling between wires-1}
\end{equation}

Together, $H_{\text{wires}}$ and $H_{\text{hop}}$ generate a 2D
Dirac dispersion in each layer, and can thus be used to model our
superlattice system. As demonstrated in Ref. \cite{Mross2016}, we
can explicitly define non-local dual fermion fields, $\tilde{\chi}_{y,z}$,
minimally coupled to a dynamical $U(1)$ gauge field, such that each
layer is described by a QED$_{3}$ of dual fermions. For simplicity,
we start by ignoring the gauge field. It will be reintroduced in the
end of the construction by imposing gauge invariance.

In terms of the dual degrees of freedom, the Hamiltonian of the decoupled
layers is given by
\begin{equation}
H_{\text{wires}}=i\tilde{v}\sum_{yz}(-1)^{y+z}\int dx\tilde{\chi}_{y,z}^{\dagger}\partial_{x}\tilde{\chi}_{y,z},\label{eq:wires-1-1}
\end{equation}

\begin{equation}
H_{\text{hop}}=t\sum_{yz}(-1)^{y+z}\int dx\left(\tilde{\chi}_{y,z}^{\dagger}\tilde{\chi}_{y+1,z}+\text{H.c.}\right).\label{coupling between wires-1-1}
\end{equation}
The non-local nature of the dual fermions does not allow us to write
inter-layer tunneling.

To simplify the derivation, it is expedient to artificially enlarge
the unit cell \cite{Mross2015,Mross2016} such that each mode in the
original layers is replaced by three chiral fermion modes $\tilde{\gamma}_{y,z,-},\tilde{\chi}_{y,z}$,$\tilde{\gamma}_{y,z,+}$
(see Fig.~\ref{fig:CDL}). The corresponding Hamiltonian is
\begin{align*}
H_{\text{hop}} & =\sum_{yz}(-1)^{y+z}\int dx\left[-\tilde{\gamma}_{y,z,-}^{\dagger}\tilde{\chi}_{y,z}+\tilde{\chi}_{y,z}^{\dagger}\tilde{\gamma}_{y,z,+}-\tilde{\gamma}_{y,z,+}^{\dagger}\tilde{\gamma}_{y+1,z,-}+\text{H.c.}\right].
\end{align*}

As we now demonstrate, interaction terms for the $\tilde{\gamma}$-fields
can induce coherence between adjacent layers, which in turn generates
inter-layer tunneling of the $\tilde{\chi}$-fields.

\subsection{Inter-layer coherence }

To make adjacent layers coherent, we introduce a local interaction
of the form
\[
H_{\text{coh}}=\sum_{_{\text{odd z}}}H_{\text{coh},z}^{\text{o}}+\sum_{_{\text{even z}}}H_{\text{coh},z}^{\text{e}},
\]
with
\begin{align*}
H_{\text{coh},z}^{\text{o}} & =u\sum_{y}\int dx\left[\tilde{\gamma}_{yz+}^{\dagger}\tilde{\gamma}_{y,z+1,+}\tilde{\gamma}_{y+1,z+1,+}^{\dagger}\tilde{\gamma}_{y+1,z,+}+\text{H.c.}\right],\\
H_{\text{coh},z}^{\text{e}} & =u\sum_{y}\int dx\left[\tilde{\gamma}_{yz-}^{\dagger}\tilde{\gamma}_{y,z+1,-}\tilde{\gamma}_{y+1,z+1,-}^{\dagger}\tilde{\gamma}_{y+1,z,-}+\text{H.c.}\right].
\end{align*}

We start by treating the odd $z$ case in details. To do so, we bosonize
the wire degrees of freedom by writing $\tilde{\gamma}_{yz\rho}\propto e^{-i\phi_{yz\rho}},$
in terms of which
\begin{align}
H_{\text{coh,z}}^{\text{o}} & =2u\sum_{y}\int dx\cos\left(\phi_{y,z,+}-\phi_{y,z+1,+}+\phi_{y+1,z+1,+}-\phi_{y+1,z,+}\right)\nonumber \\
 & =2u\sum_{y}\int dx\cos\left[2\Delta_{y}\tilde{\theta}_{y,z,z+1}\right],\label{eq:first cosine}
\end{align}
where $\tilde{\theta}_{y,z,z+1}=\frac{\phi_{y,z,+}-\phi_{y,z+1,+}}{2}$,
and $\Delta_{y}$ is the discrete derivative defined according to
$\Delta_{y}\tilde{\theta}_{y,z,z+1}=\tilde{\theta}_{y+1,z,z+1}-\tilde{\theta}_{y,z,z+1}$.

If the terms in Eq.~\eqref{eq:first cosine} flow to strong coupling,
their arguments can be assumed to weakly fluctuate around the minimum
of the cosine. Expanding in these small fluctuations, we obtain a
quadratic term of the form $u\sum_{y}(\Delta_{y}\tilde{\theta}_{y,z,z+1})^{2}$.
Taking this term together with the quadratic Luttinger-liquid Hamiltonian,
the operator $ e^{2i\tilde{\theta}_{y,z,z+1}}$
acquires an expectation value. We thus obtain a spontaneous breaking of the $U(1)$
symmetry $\tilde{\theta}\rightarrow\tilde{\theta}+\text{const}$.
Therefore, at low energies, one can write $\tilde{\theta}_{y,z,z+1}=\tilde{\theta}_{0}+\frac{1}{2}a_{yz}^{3}(x)$,
where $a_{yz}^{3}$ is a slowly varying Goldstone mode.

For even $z$, we repeat the same analysis in terms of the $\tilde{\gamma}_{-}$-fields.
For these layers, we define $\tilde{\theta}_{y,z,z+1}=\frac{\phi_{y,z,-}-\phi_{y,z+1,-}}{2}$,
and find
\begin{align}
H_{\text{coh,z}}^{\text{e}} & =2u\sum_{y}\int dx\cos\left[2\Delta_{y}\tilde{\theta}_{y,z,z+1}\right].\label{eq:second cosine}
\end{align}
Following the same arguments as for the odd $z$ case, we again achieve
spontaneous symmetry breaking.

Once all the layers are coherent, we generate inter-layer hopping
terms for $\tilde{\chi}$. To do so, we write local interaction terms
involving correlated tunneling in adjacent layers, as shown in Fig.
\ref{fig:CDLcoherence}:

\begin{equation}
H_{z}=\sum_{y,z}\int dx\begin{cases}
\tilde{\chi}_{y,z}^{\dagger}\tilde{\chi}_{y,z+1}\tilde{\gamma}_{y,z+1,+}^{\dagger}\tilde{\gamma}_{y,z,+}+\text{H.c.} & \text{odd }z\\
\tilde{\chi}_{y,z}^{\dagger}\tilde{\chi}_{y,z+1}\tilde{\gamma}_{y,z+1,-}^{\dagger}\tilde{\gamma}_{y,z,-}+\text{H.c.} & \text{even }z
\end{cases}.\label{eq:Hz}
\end{equation}
Notice that these terms do not involve direct tunneling of dual fermions
between different layers. To effectively generate such tunneling,
we use the inter-layer coherence and write
\begin{align*}
\tilde{\gamma}_{y,z+1,+}^{\dagger}\tilde{\gamma}_{y,z,+} & \sim e^{-2i\tilde{\theta}_{y,z,z+1}}\sim e^{-ia_{yz}^{3}(x)}\text{ for odd }z\\
\tilde{\gamma}_{y,z+1,-}^{\dagger}\tilde{\gamma}_{y,z,-} & \sim e^{-2i\tilde{\theta}_{y,z,z+1}}\sim e^{-ia_{yz}^{3}(x)}\text{ for even }z.
\end{align*}
Therefore, $H_{z}$ takes the form
\[
H_{z}=\sum_{y,z}\int dx\left[\tilde{\chi}_{yz}^{\dagger}\tilde{\chi}_{yz+1}e^{-ia_{yz}^{3}(x)}+\text{H.c.}\right].
\]

Thus, despite the non-local nature of the dual fermions, they can
tunnel between the layers through the phenomenon of inter-layer coherence.

As noted in the main text, the gauge degrees of freedom can be reintroduced
by enforcing gauge invariance. However, the inter-layer tunneling terms
reduce the gauge symmetry from an independent $U(1)$ symmetry for
each interface to a single $U(1)$ symmetry for the entire three dimensional
system. This promotes the gauge degrees of freedom to form a truly
$3+1$ dimensional $U(1)$ gauge theory, with the Goldstone mode $a^{3}$
acting as the fourth component of $a$.

In the next section, we demonstrate that the inter-layer tunneling
terms generate a $3+1$ dimensional Dirac theory, leading to the $\text{QED}_{4}$
parent state.

\section{Derivation of the $\text{QED}_{4}$ fixed point}

\label{sec:Appendix 1}

In this section, we demonstrate the emergence of the $\text{QED}_{4}$
fixed point. In the absence of coupling between the layers, the
action takes the form
\begin{equation}
{\cal S}_{\text{dual}}=\sum_{z}\int d^{2}xdt\left[i\bar{\tilde{\Psi}}_{z}\gamma_{\mu z}\left(\partial^{\mu}-ia_{z}^{\mu}\right)\tilde{\Psi}_{z}+\mathcal{L}_{\text{Maxwell}}\left[a_{z}^{\mu}\right]\right]\label{eq: dual action supp}
\end{equation}
in the dual formulation, where $\mathcal{L}_{\text{Maxwell}}$ denotes
the $2+1$-dimensional Maxwell term in each layer.

Upon inducing inter-layer coherence, we effectively generate tunneling
terms of the form
\begin{equation}
{\cal S}_{\text{int}}=\chi_{0}\sum_{z}\sum_{\alpha=\uparrow,\downarrow}\int d^{2}xdt\left(e^{-ia_{z}^{3}}\tilde{\psi}_{z,\alpha}^{\dagger}\tilde{\psi}_{z+1,\alpha}+\text{H.c.}\right).\label{eq: HS mean field supp}
\end{equation}

Next, we note that in the absence of gauge fluctuations the inter-layer
action [Eq.~\eqref{eq: HS mean field supp}] is associated with the
spectrum $E=2\chi_{0}\cos k_{z}$, which gives rise to low-energy
fermion excitations around $k_{z}=\pm\frac{\pi}{2}$. Focusing on these low energy degrees of freedom, we write
\begin{equation}
\tilde{\Psi}_{z}=\tilde{\Psi}_{z,+}e^{i\frac{\pi z}{2}}+\tilde{\Psi}_{z,-}e^{-i\frac{\pi z}{2}}.\label{eq:low energy}
\end{equation}
Notice that in this mean field approximation the unit cell consists of two layers, meaning the Brillouin zone is defined for $k_z$ between $-\pi/2$ and  $\pi/2$, and the above low energy degrees of freedom coincide at the edge of the Brillouin zone.

 Inserting this into Eqs.~\eqref{eq: dual action supp} and \eqref{eq: HS mean field supp}
and dropping the spatially oscillating terms, which do not affect
the low energy description, we obtain
\begin{align}
{\cal S}= & \sum_{z}\int d^{2}xdt\left\{ \sum_{\rho=\pm}\left[\left(i\chi_{0}\rho e^{-ia_{z}^{3}}\tilde{\Psi}_{z,\rho}^{\dagger}\tilde{\Psi}_{z+1,\rho}+\text{H.c.}\right)\right.\right.\nonumber \\
 & \left.+i\tilde{\Psi}_{z\rho}^{\dagger}\left(\partial^{0}-ia_{z}^{0}\right)\tilde{\Psi}_{z\rho}\right]\nonumber \\
 & +i\tilde{\Psi}_{z+}^{\dagger}\sigma_{3}\left(\partial^{1}-ia_{z}^{1}\right)\tilde{\Psi}_{z-}+\text{H.c.}\nonumber \\
 & \left.+i\tilde{\Psi}_{z+}^{\dagger}\sigma_{2}\left(\partial^{2}-ia_{z}^{2}\right)\tilde{\Psi}_{z-}+\text{H.c.}\right\} .\label{eq:S before continuum}
\end{align}
Defining the four-component spinor $\Psi'=\begin{pmatrix}\Psi_{\uparrow+} & \Psi_{\uparrow-} & \Psi_{\downarrow+} & \Psi_{\downarrow-}\end{pmatrix}$,
we take the natural continuum limit and obtain
\begin{align}
{\cal S} & =i\int d^{3}x\bar{dt\Psi}'\Gamma_{\mu}\left(\partial^{\mu}-ia^{\mu}\right)\Psi',\label{eq:Dirac action Supp}
\end{align}
where $\Gamma_{0}=\sigma_{1}\tau_{1},\Gamma_{1}=-i\sigma_{2},\Gamma_{2}=i\sigma_{3},\Gamma_{3}=-i\sigma_{1}\tau_{2}$,
and the $\tau_{i}$-matrices are the Pauli matrices acting on the
$\rho=\pm$ space. It can easily be verified that the $\Gamma$-matrices
satisfy the Clifford algebra, indicating that the action in Eq.~\eqref{eq:Dirac action Supp}
indeed describes Dirac fermions coupled to a $U(1)$ gauge field in
$3+1$-dimensions. Notice that in moving from Eq.~\eqref{eq:S before continuum}
to Eq.~\eqref{eq:Dirac action Supp}, we have rescaled spacetime to
make the Dirac action appear isotropic.

The kinetic action of the dynamical gauge field $a_{\mu}$ generically
consists of all gauge invariant terms. The most relevant among these
is the Maxwell term $\mathcal{L}_{\text{Maxwell}}=-\frac{1}{4e^{2}}F_{\mu\nu}F^{\mu\nu}$
with $\mu,\nu=0,\cdots,3$. Notice that due to the anisotropic nature
of our realization, we generally expect an anisotropic version
of the Maxwell action.

\section{The composite Weyl semimetal phase}

With the addition of the mass term

\begin{equation}
{\cal S}_{m}=m\sum_{z}\int d^{2}xdt\left(\tilde{\psi}_{z,\uparrow}^{\dagger}\tilde{\psi}_{z,\downarrow}+H.c.\right),\label{eq:time reversal breaking Supp}
\end{equation}
it can easily be checked that the spectrum generated by Eqs.~\eqref{eq: HS mean field supp}
and \eqref{eq:time reversal breaking Supp} hosts low-energy excitations
around four points of the form $\vec{k} = (0,0,k_z)$. Defining $k_{0}=\cos^{-1}\left(\frac{\left|m\right|}{2\chi_{0}}\right)$, two of these points are located at $k_z = \pm k_0$ and associated with the wavefunction $\frac{1}{\sqrt{2}}\begin{pmatrix}1 & -1\end{pmatrix}^{T}$, while the other two, located at $k_z = \pm (\pi - k_0)$, are associated with $\frac{1}{\sqrt{2}}\begin{pmatrix}1 & 1\end{pmatrix}^{T}$.

We therefore write at low energies
\begin{align}
\tilde{\Psi}_{z} & =\frac{1}{\sqrt{2}}\left[\left(\tilde{\psi}_{z,++}e^{ik_{0}z}+\tilde{\psi}_{z,-+}e^{-ik_{0}z}\right)\begin{pmatrix}1\\
-1
\end{pmatrix}\right.\label{eq:low energy-1-1}\\
 & \left.+(-1)^{z}\left(\tilde{\psi}_{z,--}e^{-ik_{0}z}+\tilde{\psi}_{z,+-}e^{ik_{0}z}\right)\begin{pmatrix}1\\
1
\end{pmatrix}\right].\nonumber
\end{align}
Inserting this into the action in Eqs.~\eqref{eq: dual action supp},\eqref{eq: HS mean field supp},
and \eqref{eq:time reversal breaking Supp}, we obtain
\begin{align}
{\cal S}= & \sum_{z}\sum_{\beta=\pm}\int d^{2}xdt\left\{ \sum_{\gamma}\left[\left(i\tilde{\chi}_{0}\beta\gamma e^{-ia_{z}^{3}}\tilde{\psi}_{z,\beta\gamma}^{\dagger}\tilde{\psi}_{z+1,\beta\gamma}+H.c.\right)\right.\right.\nonumber \\
 & \left.+i\tilde{\psi}_{z\beta\gamma}^{\dagger}\left(\partial^{0}-ia_{z}^{0}\right)\tilde{\psi}_{z\beta\gamma}\right]\nonumber \\
 & +i\tilde{\psi}_{z\beta-}^{\dagger}\left(\partial^{1}-ia_{z}^{1}\right)\tilde{\psi}_{z,\beta+}+H.c.\nonumber \\
 & \left.-\tilde{\psi}_{z\beta-}^{\dagger}\left(\partial^{2}-ia_{z}^{2}\right)\tilde{\psi}_{z,\beta+}+H.c.\right\} ,\label{eq:S before continuum-1}
\end{align}
with $\tilde{\chi}_{0}=\sqrt{\chi_{0}^{2}-\frac{m^{2}}{4}}$.

Taking the continuum limit by assuming $\tilde{\psi}_{z,\beta\gamma}$
to vary slowly in the $z$ direction, we get the low-energy action

\begin{align*}
{\cal S}=i\sum_{\beta=\pm1} & \int d^{3}xdt\tilde{\Psi}_{\beta}^{\dagger}\sigma_{\mu}^{\beta}\left(\partial^{\mu}-ia_{j}^{\mu}\right)\tilde{\Psi}_{\beta},
\end{align*}
where $\sigma_{\mu}$ are the Pauli matrices acting on the $\gamma$-indices,
$\sigma_{\mu}^{\beta}=\sigma_{\mu}$ for $\mu=0,1,2$ and $\sigma_{3}^{\beta}=\beta\sigma_{3}$.
The resulting theory indeed describes two separate Weyl theories coupled
to a gauge field. As pointed out in the main text, $a$ generally acquires a mass through the Higgs mechanism, thus leaving only Weyl fermions at low energies.
Notice that while these two Weyl theories are technically
equivalent to a single Dirac theory, they are associated with degrees
of freedom located at different points in the Brillouin zone, prompting
us to regard them as separate theories.

\section{The monopole propagator}

In Sec.~\ref{sec:Appendix 1}, we obtained an effective $\text{QED}_{4}$
theory, with the action
\begin{equation}
{\cal S}_{\text{QED}_{4}}=\int d^{4}x\left(-\frac{1}{4e^{2}}F_{\mu\nu}F^{\mu\nu}+i\bar{\tilde{\Psi}}\gamma^{\mu}\left(\partial_{\mu}-ia_{\mu}\right)\tilde{\Psi}\right).\label{eq:QED4}
\end{equation}
In what follows, we calculate the monopole propagator within this theory,
\begin{equation}
C(\vec{x},t)=\left\langle M(\vec{x}_{t},t)M^{\dagger}(\vec{x}_{0},0)\right\rangle ,\label{eq:monopole propagator}
\end{equation}
where $M^{\dagger}$ is an operator that creates a $4\pi$ monopole.
To calculate this propagator, we first employ the asymptotic freedom
of $\text{QED}_{4}$ at low energies to neglect the coupling to the
dual Dirac theory. This allows us calculate Eq.~\eqref{eq:monopole propagator}
within the free Maxwell theory (i.e., the first term of Eq.~\eqref{eq:QED4}).
Second, we use the electromagnetic duality of the Maxwell theory to
the calculate the propagator of the electronic charge instead, and
argue that the result should be identical in the magnetic case.

Within the Maxwell theory, the charge propagator can be written as
a path integral
\[
C(\vec{x}_{\mathrm{f}}-\vec{x}_{\mathrm{i}},t_{\mathrm{f}}-t_{\mathrm{i}})=\int_{\vec{x}_{\mathrm{i}}(t_{\mathrm{i}})}^{\vec{x}_{\mathrm{f}}(t_{\mathrm{f}})}{\cal{D}}\vec{x}\int {\cal D}a_{\mu}e^{i{\cal S}(x,a)},
\]
where $S=\int\left[-\frac{1}{4e^{2}}F_{\mu\nu}F^{\mu\nu}+a_{\mu}J^{\mu}\right],$
with $J^{\mu}=qv^{\mu}\delta(\vec{x}-\vec{x}(t))$ and $v^{\mu}=\frac{dx^{\mu}}{dt}=\left(1,d\vec{x}/dt\right)$.
We integrate over all the paths starting at $\vec{x}_{\mathrm{i}}$ and ending
at $\vec{x}_{\mathrm{f}}$, with the dynamics determined only by the coupling
to the electromagnetic field.

Defining the photon propagator $G_{\mu\nu}(x)$, we integrate out
the electromagnetic field and obtain the effective action
\[
C(\vec{x}_{\mathrm{f}}-\vec{x}_{\mathrm{i}},t_{\mathrm{f}}-t_{\mathrm{i}})\equiv\int_{\vec{x}_{\mathrm{i}}(t_{\mathrm{i}})}^{\vec{x}_{\mathrm{f}}(t_{\mathrm{f}})}{\cal D}\vec{x}e^{\mathrm{\mathbb{\mathcal{S}}}_{\mathrm{eff}}[x]},
\]
where
\begin{align*}
\mathrm{\mathbb{\mathcal{S}}}_{\mathrm{eff}}[x] & =q^{2}\int dtdt'v^{\mu}(t)v^{\nu}(t')G^{\mu\nu}\left[\vec{x}(t)-\vec{x}(t'),t-t'\right].
\end{align*}
Adopting the Feynman gauge, we have
\[
G_{\mu\nu}(x)\propto\frac{e^{2}\eta_{\mu\nu}}{x^{2}},
\]
with $x$  a four-vector. Plugging this in, we obtain
\begin{align}
\mathrm{\mathbb{\mathcal{S}}}_{\mathrm{eff}} & =\frac{\left(qe\right)^{2}}{8\pi^{2}}\int dtdt'v^{\mu}(t)v^{\nu}(t')\eta_{\mu\nu}\frac{1}{\left(t-t'\right)^{2}-\left(\vec{x}(t)-\vec{x}(t')\right)^{2}}\label{eq:seff}\\
 & =\frac{\left(qe\right)^{2}}{8\pi^{2}}\int dx^{\mu}dx'^{\nu}\eta_{\mu\nu}\frac{1}{\left(x-x'\right)^{2}},
\end{align}
with $x,x'$ four-vectors.

The effective action can now be cast as
\begin{equation}
\mathrm{\mathbb{\mathcal{S}}}_{\mathrm{eff}}=\frac{\left(qe\right)^{2}}{8\pi^{2}}\int d\tau d\tau'\frac{dx^{\mu}}{d\tau}\frac{dx'^{\nu}}{d\tau'}\eta_{\mu\nu}\frac{1}{\left(x(\tau)-x(\tau')\right)^{2}},\label{eq:covariant form}
\end{equation}
in terms of the (Lorentz invariant) proper time $\tau$ along the
path. We now split the integral over $\tau$ into infinitely many
infinitesimal segments, each of which is Lorentz invariant. Therefore,
for each $\tau$ we can choose to write the corresponding segment
in an inertial frame of reference comoving with $x^{\mu}(\tau)$
(Notice that for each point $\tau$ on the path we use a distinct
frame of reference). In this case, $\frac{dx^{\mu}}{d\tau}=(1,0,0,0)$,
and we get
\[
\frac{dx^{\mu}}{d\tau}\frac{dx'^{\nu}}{d\tau'}\eta_{\mu\nu}=\frac{dt'}{d\tau'}.
\]

As can be seen by examining the denominator of Eq.~\eqref{eq:covariant form},
the dominant contributions arise from the vicinity of $\tau'=\tau$.
Therefore, to leading order in $\tau-\tau'$ we take
\begin{align*}
x^{\mu}(\tau)-x^{\mu}(\tau') & =\frac{dx^{\mu}(\tau)}{d\tau}\left(\tau'-\tau\right)=\delta_{0}^{\mu}\left(\tau-\tau'\right)\\
dt' & =d\tau'=d\tau.
\end{align*}
This results in
\[
\mathrm{\mathbb{\mathcal{S}}}_{\mathrm{eff}}=\frac{\left(qe\right)^{2}}{8\pi^{2}}\int_{0}^{T}d\tau\int_{0}^{T}d\tau'\frac{1}{\left(\tau-\tau'\right)^{2}},
\]
with $T$ the proper time at the end of the path. To regularize
the integral we introduce a cutoff:
\begin{align*}
\mathrm{\mathbb{\mathcal{S}}}_{\mathrm{eff}} & =\frac{\left(qe\right)^{2}}{8\pi^{2}}\int_{0}^{T}d\tau\int_{0}^{T}d\tau'\frac{1}{\left(\tau-\tau'+i\epsilon\right)^{2}}\\
 & =\frac{\left(qe\right)^{2}}{8\pi^{2}}\left[2\log\frac{\epsilon}{T}-\log\left(1+\left(\frac{\epsilon}{T}\right)^{2}\right)\right]\approx\frac{\left(qe\right)^{2}}{4\pi^{2}}\log\frac{\epsilon}{T}.
\end{align*}
This finally gives us the propagator
\[
C(x_{\mathrm{f}}-x_{\mathrm{i}})\propto e^{\frac{\left(qe\right)^{2}}{4\pi^{2}}\log\frac{\epsilon}{\left|x_{\mathrm{f}}-x_{\mathrm{i}}\right|}}\propto\left(\left|x_{\mathrm{f}}-x_{\mathrm{i}}\right|\right)^{-\frac{\left(qe\right)^{2}}{4\pi^{2}}}.
\]

\section{Distinction between gapped phases}

In this section we comment on the distinction between the two gapped
phases discussed in the main text (in the absence of additional discrete
symmetries).

As discussed in the main text, for $m=\Delta=0$, one finds two dual
Weyl nodes located at the edge of the Brillouin zone at $k_{z}=\pm\frac{\pi}{2}$
(and $k_{x}=k_{y}=0$), thus forming the parent Dirac theory.

Introducing $m,\Delta\neq0$, the nodes are shifted to $k_{z}=\pm k_{0}$,
with
\begin{equation}
k_{0}=\cos^{-1}\left(\frac{\sqrt{m^{2}-\Delta^{2}}}{2\chi_{0}}\right).\label{eq:location of cones}
\end{equation}
It is evident that $m$ pushes the nodes toward the origin at $k_{z}=0$,
while $\Delta$ pushes them toward the edge of the Brillouin zone.
If the cones meet at either of these points, they generically annihilate
each other, leading to completely gapped phases. Depending on where
the cones meet, we obtain two distinct gapped phases.

One can understand the distinction between the two phases by examining
the Bloch Hamiltonian at fixed values of $k_{z}$: $H_{k_{z}}(k_{x},k_{y})$.
Given $k_{z}$, these two-dimensional Hamiltonians can be associated
with an integer Chern number $C(k_{z})$. Notice that the Chern number
is defined in the BDG formulation, and therefore counts the number
of chiral Majorana edge modes for the effective two-dimensional models.
In particular, $C=2$ corresponds to a single chiral fermion on the
edge (i.e., two Majorana modes).

In our model, this Chern number is given by
\[
C(k_{z})=\begin{cases}
0 & \left|k_{z}\right|<k_{0}\\
2 & \left|k_{z}\right|>k_{0}
\end{cases}.
\]

Indeed, the points where the Chern number transitions from one value
to another and the gap must therefore close, correspond to the locations
of the Weyl nodes.

From Eq.~\eqref{eq:location of cones}, we see that the Weyl cones survive
as long as $\left|\Delta\right|<\left|m\right|$ and $m^{2}<4\chi_{0}^{2}+\Delta^{2}$.
If $\Delta$ is increased such that $\left|\Delta\right|=\left|m\right|$,
the nodes meet at the edge of the Brillouin zone. As they move toward
the edge of the Brillouin zone, the region of non-zero $C$ shrinks,
and eventually vanishes, leaving us with $C=0$ throughout the Brillouin
zone when they meet. For $\left|\Delta\right|>\left|m\right|$, the
two cones gap out, leading to a trivially gapped state with $C=0$
for all $k_{z}$.

On the other hand, if we increase $m$ instead of $\Delta$, the nodes
are shifted toward the origin of the Brillouin zone, thus enlarging
the non-trivial region of $C=2$. Eventually, at $m^{2}=4\chi_{0}^{2}+\Delta^{2}$,
the two cones meet at the origin, and the non-trivial region of $C=2$
covers the entire Brillouin zone. For larger values of $m$, the cones
annihilate out and we end up with a gapped phase associated with a
Chern number $C=2$ for each $k_{z}$. Such a phase is adiabatically
connectable to a stack of Chern insulators (with a single chiral fermionic
edge mode) forming a 3D quantum Hall state of dual fermions. While such a state has no electronic response, it possesses a thermal Hall conductivity, and is thus referred to as a thermal Hall insulator.

As we discussed in the main text, additional scenarios arise where
$\Delta$ is not identical in all layers. Generically, each of the
Weyl cones splits into two Majorana-Weyl cones \cite{Meng2012}, located
at $k_{z}=-k_{2},-k_{1},k_{1},k_{2}$. In this case, the Chern number
is given by
\[
C(k_{z})=\begin{cases}
0 & \left|k_{z}\right|<k_{1}\\
1 & k_{1}<\left|k_{z}\right|<k_{2}\\
2 & \left|k_{z}\right|>k_{2}
\end{cases}.
\]
A distinct gapped phase can be stabilized if $k_{2}$ is shifted toward
the edge of the Brillouin zone, while $k_{1}$ is shifted toward the
origin. This results in a gapped phase associated with $C=1$ for
all $k_{z}$. Such a phase is adiabatically connectable to a stack
of dual-fermion $p+ip$ superconductors associated with each two-layer unit cell. In terms of the electrons, while the case $C=2$ can be understood as a stack of Abelian states, the $C=1$ state necessarily requires non-Abelian states.

\section{Recovering non-interacting gapped states}

In this section, we address the subtleties associated with recovering
trivial non-interacting phases starting from the dual formulation.
To be specific, we will focus on the phases that arise when the interfaces
are gapped by time-reversal breaking mass terms.

In this case, the action of the non-interacting electrons is given
by
\begin{align*}
{\cal S} & =\sum_{z}i\int d^{2}xdt\bar{\Psi}_{z}\gamma_{\mu z}\left(\partial^{\mu}-iA_{z}^{\mu}\right)\Psi_{j}\\
 & +\sum_{z}m_{z}\int d^{2}xdt\bar{\Psi}_{z}\Psi_{z}.
\end{align*}
Depending on the sign of $m_{z}$, each interface hosts a massive
Dirac fermion, known to induce a Hall conductance of $\sigma_{xy}=\pm1/2$.
Therefore, integrating out the electrons, we obtain an effective Chern-Simons
action for the external electromagnetic field
\begin{align}
{\cal S} & =\frac{1}{8\pi}\sum_{z}\text{sgn}\left(m_{z}\right)\epsilon^{\mu\nu\kappa}A_{\mu,z}\partial_{\nu}A_{\kappa,z}.\label{eq:CS for A}
\end{align}

We now turn to derive the effective action for the probe field starting
from the dual formulation, where the full action is given by
\begin{align*}
{\cal S}_{\text{dual}} & =\sum_{z}\int d^{2}xdt\left[i\bar{\tilde{\Psi}}_{z}\gamma_{\mu z}\left(\partial^{\mu}-ia_{z}^{\mu}\right)\tilde{\Psi}_{j}+\frac{1}{4\pi}a_{z}^{\mu}\epsilon_{\mu\nu\kappa}\partial_{\nu}A_{\kappa,z}+\mathcal{L}_{\text{Maxwell}}\left[a_{z}^{\mu}\right]\right]\\
 & +\sum_{z}m_{z}\int d^{2}xdt\bar{\tilde{\Psi}}_{z}\tilde{\Psi}_{z}.
\end{align*}
Notice that we have used the identity $\bar{\tilde{\Psi}}_{z}\tilde{\Psi}_{z}=\bar{\Psi}_{z}\Psi_{z}$.
Integrating out the dual fermions, we generate a Chern-Simons action
for the emergent gauge field $a$
\begin{align}
{\cal S}_{\text{dual}} & =\sum_{z}\int d^{2}xdt\left[\frac{1}{8\pi}\text{sgn}\left(m_{z}\right)\epsilon^{\mu\nu\kappa}a_{\mu}\partial_{\nu}a_{\kappa}+\frac{1}{4\pi}a_{z}^{\mu}\epsilon_{\mu\nu\kappa}\partial_{\nu}A_{\kappa,z}+\mathcal{L}_{\text{Maxwell}}\left[a_{z}^{\mu}\right]\right].\label{eq:CS for a}
\end{align}
A particularly subtle situation arises when $\text{sgn}\left(m_{z}\right)=(-1)^{z}$.
In this case, Eq.~\eqref{eq:CS for a} becomes
\begin{align}
{\cal S}_{\text{dual}} & =\sum_{z}\int d^{2}xdt\left[\frac{1}{8\pi}(-1)^{z}\epsilon^{\mu\nu\kappa}a_{\mu}\partial_{\nu}a_{\kappa}+\frac{1}{4\pi}a_{z}^{\mu}\epsilon_{\mu\nu\kappa}\partial_{\nu}A_{\kappa,z}+\mathcal{L}_{\text{Maxwell}}\left[a_{z}^{\mu}\right]\right].\label{eq:CS for a, oscillating}
\end{align}
It seems natural to take the continuum limit in $z$, and neglect
the first term due to its oscillating nature. However, if we then
integrate out the $a$-field, we obtain a mass term for the electromagnetic
field, which would indicate superconductivity on the various interfaces.
Recalling that the configuration we are describing realizes non-interacting
insulating phases, the above continuum limit is clearly incorrect.
This highlights a subtlety that may arise in using the dual formulation.

To recover the correct behavior, we return to Eq.~\eqref{eq:CS for a, oscillating}
and integrate out $a$ \textit{before} taking the continuum limit.
Indeed, the lowest order term in the resulting action for the probe
field is the Chern-Simons term given in Eq.~\eqref{eq:CS for A}.

\end{document}